\documentclass[showpacs,prl,twocolumn]{revtex4}
\usepackage{graphicx}
\usepackage{enumerate}
\usepackage{hyperref}
\usepackage{textcomp}
\usepackage{amsmath}
\usepackage{amssymb}
\usepackage{pgf}
\usepackage{diagbox}
\usepackage{booktabs}
\usepackage{bbold}
\usepackage{subfigure}
\usepackage{epstopdf}
\usepackage{color}
\usepackage{soul}
\newcommand{\C}{{\hat c}}
\newcommand{\D}{{\hat d}}
\newcommand{\Hamm}{{ \mathcal{\hat H}}}
\newcommand{\I}{{ \mathcal{\hat I}}}

\newcommand{\T}{{ \mathcal{\hat T}}}
\newcommand{\Ham}{{\hat H}}

\begin{document}

\title{Topological phases in odd-legs frustrated synthetic ladders}
\author{Simone Barbarino$^{1}$\footnote{simone.barbarino@sns.it}, Marcello Dalmonte$^{3}$, Rosario Fazio$^{3,4}$, Giuseppe E. Santoro$^{1,2,3}$}

\affiliation{
$^1$ SISSA, Via Bonomea 265, I-34136 Trieste, Italy\\
$^2$ CNR-IOM Democritos National Simulation Center, Via Bonomea 265, I-34136 Trieste, Italy\\
$^3$ International Centre for Theoretical Physics (ICTP), P.O. Box 586, I-34014 Trieste, Italy\\
$^4$ NEST, Scuola Normale Superiore $\&$ Istituto Nanoscienze-CNR, I-56126 Pisa, Italy 
}

\begin{abstract}

The realization of the Hofstadter model in a strongly anisotropic ladder geometry has now become possible in one-dimensional optical lattices with a synthetic dimension. 
In this work, we show how the Hofstadter Hamiltonian in such ladder configurations hosts a topological phase of matter which is radically different from its two-dimensional counterpart. This topological phase stems directly from the hybrid nature of the ladder geometry, and is protected by a properly defined inversion symmetry.
We start our analysis considering the paradigmatic case of a three-leg ladder which supports a topological phase exhibiting the typical features of topological states in one dimension: robust fermionic edge modes, a degenerate entanglement spectrum and a non-zero Zak phase; then, we generalize our findings -  addressable in the state-of-the-art cold atom experiments - to ladders with an higher number of legs.

\end{abstract}

\maketitle

\begin{figure}
	\begin{center}
  	\includegraphics[width=\columnwidth]{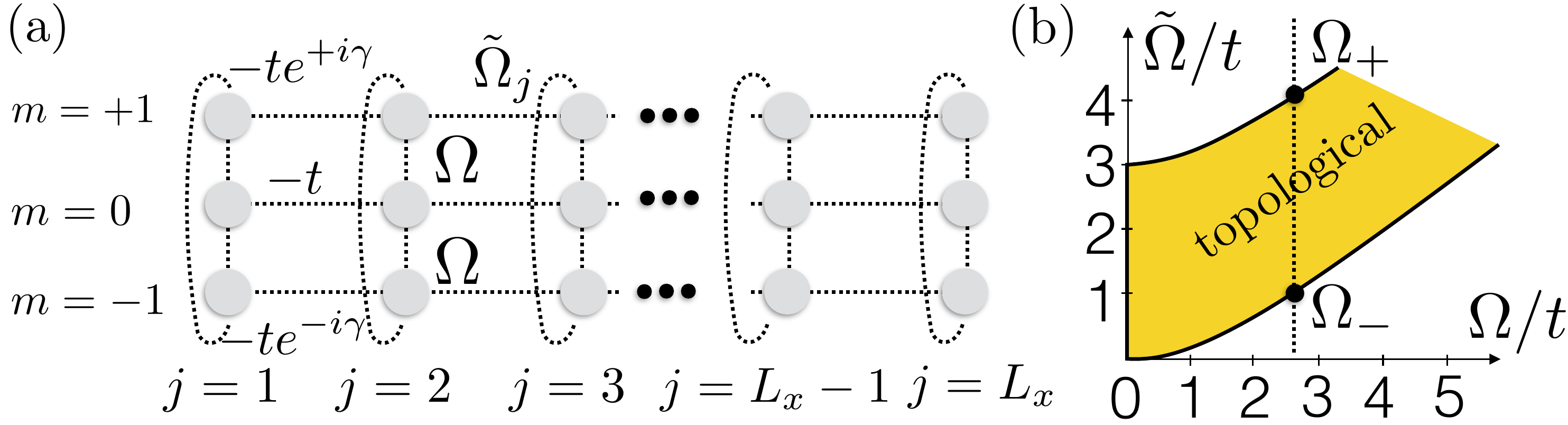}
	\end{center}
	\caption{(a) Schematic representation of the Hofstadter Hamiltonian on a three-leg ladder; in the presence of a non-vanishing coupling between the extremal states $m=+1$ and $m=-1$ ($\tilde \Omega_j \neq 0$), the ladder is equivalent to a cylinder. (b) Phase diagram for a three leg-ladder.}
	\label{fig:model}
\end{figure}

The Hofstadter problem~\cite{Hofstadter75} describing a particle hopping on a two-dimensional lattice pierced by a magnetic field, is a paradigm of quantum 
mechanics. Formulated more then forty years ago, it embeds a multitude of seminal notions in modern condensed matter physics~\cite{Bernevig13}: topological bands, edge
excitations, fractal properties of the spectrum, just to mention some of them. Despite its apparent simplicity and the enormous body of investigation both theoretical and experimental~\cite{Dean13, Aidelsburger13, Miyake13, Aidelsburger15, Atala14}, the Hofstadter problem still hides some surprises, as we are going to discuss in the following.  

The motivation of our work stems from the recent realization~\cite{Mancini15,Stuhl15,Kolkowitz17} of the Hofstadter Hamiltonian in optical lattices with a synthetic 
dimension.  The possibility of engineering an additional (synthetic) few sites long dimension~\cite{Boada12,Celi14}
with non-trivial boundary conditions by using some internal degrees of freedom of the atoms has encouraged  the study of the Hofstadter Hamiltonian in a strongly 
anisotropic geometry~\cite{Mugel17}. 

Is this a new territory for the Hofstadter problem or are we bound to detect a smooth crossover from a two- 
to a one-dimensional behavior?  A na\"ive expectation would induce to think that synthetic lattices can simulate a two-dimensional geometry when the transverse 
dimension is much longer than the correlation length of the system along the transverse dimension itself, and they turn out to be effectively one-dimensional 
when the number of sites in the transverse direction is small, such as in synthetic ladders. In this work, we show that in the presence of a selected 
applied magnetic field and for an odd number of legs ($L_y$), this simple picture fails and the Hofstadter Hamiltonian can host unexpected  topological phases  for suitable values of the flux piercing the ladder.
These phases, addressable in the state-of-the-art cold-atom experiments~\cite{Mancini15,Stuhl15}, are protected by a hidden inversion symmetry and exhibit the typical features of topological states in one dimension, \textit{i.e.}, exponentially localized degenerate edge states, a non-zero Zak phase~\cite{Zak89,Xiao10,Mazza15}, and a degenerate entanglement 
spectrum~\cite{Pollmann10,Fidkowski10}.  Note, however, that they can be classified neither as two-dimensional since the synthetic dimension is too short, 
nor as one-dimensional since they are strongly dependent on the choice of the boundary conditions along the synthetic dimension. In the light of this, 
the experimental realization of these topological phases is intimately related to the existence of non-vanishing boundary conditions along the synthetic dimension which is indeed a peculiarity of synthetic lattices. 

In the past, great efforts have been devoted  to searching topological states stabilized by the Hofstadter Hamiltonian in a ladder geometry. 
For example, the Hofstadter Hamiltonian on a two-leg ladder can give rise to a topological phase provided an additional \textit{ad-hoc} off-diagonal 
term is added~\cite{Hugel14}, and similar topological phases are also predicted in the same ladder geometry when the magnetic flux per plaquette  
is spatial oscillating~\cite{Grusdt14,Sun16}.  While these models go in the direction that one can have topological phases protected by the 
inversion symmetry by introducing a spatial modulation of the physical parameters~\cite{Chen12,Chiu13,Shiozaki14,Guo15}, the phases discussed here 
represent, to the best of our knowledge, the first example of non-interacting topological states stabilized by the Hofstadter Hamiltonian  in a ladder geometry with synthetic length $L_y>2$. 

\paragraph*{Model.} We consider a one-dimensional chain of length $L_x$ with fermionic atoms endowed with $2I+1$ internal spin states described by the Hamiltonian  $\hat H= \hat H_x + \hat H_y$ with
\begin{align}
	&\Ham_x= \sum_{j=1}^{L_x} \sum_{m=-I}^{+I} \left( -t \, e^{i \gamma m}\, \D^\dagger_{j+1,m} \D_{j,m} + \mathrm{H.c.} \right) \nonumber\\
	&\Ham_y= \sum_{j=1}^{L_x} \left( \sum_{m=-I}^{+I-1}  \Omega\,  \D^\dagger_{j,m} \D_{j,m+1} + \tilde\Omega_j \, \D_{j,+I}^\dagger \D_{j,-I} + \mathrm{H.c.}  \right)\, ,
	\label{ham_hofstadter_soc}
\end{align}
where the canonical operator $\D^{(\dagger)}_{j,m}$ annihilates (creates) a fermion at site $j=1,\dots, L_x$ with internal state $m=-I,\dots, I$; $I$  integer (half-integer) corresponds to an odd (even) number of internal states. 

The Hamiltonian $\hat H$ can be interpreted as the Hofstadter  Hamiltonian for spinless fermions on a $L_y=(2I+1)$-leg ladder. The tunneling amplitudes $-t \,e^{i \gamma m}$ between neighboring sites along the real direction generate the magnetic field corresponding to a uniform  flux $+\gamma$ per plaquette (in the following we set $t=1$);  
$\Omega$ is the coupling strength between adjacent spin states along the synthetic direction, while the coupling between the extremal states $-I$ and $+I$ is $\tilde \Omega_j \equiv \tilde \Omega \, e^{-i\gamma (2I+1) j}$, see Fig.~\ref{fig:model}(a). This choice of $\tilde \Omega_j$ guarantees that the flux in each possible plaquette is $+\gamma$, see Supplementary Material. 
We stress here that $\hat H$ is not endowed with anti-unitary symmetries since time-reversal, particle-hole, and chiral symmetries are broken. For this reason, 
$\hat H$ belongs to the unitary symmetry class A of the Altland Zirnbauer classification~\cite{Altland97,Schnyder08,Ludwig15,Chiu16}.
In the following, we show that topological phases appear when $L_y$ is odd and $\gamma=2\pi/L_y$.

\paragraph*{Topological phase for $L_y=3$.} 

\begin{figure}
	\begin{center}
  	\includegraphics[width=\columnwidth]{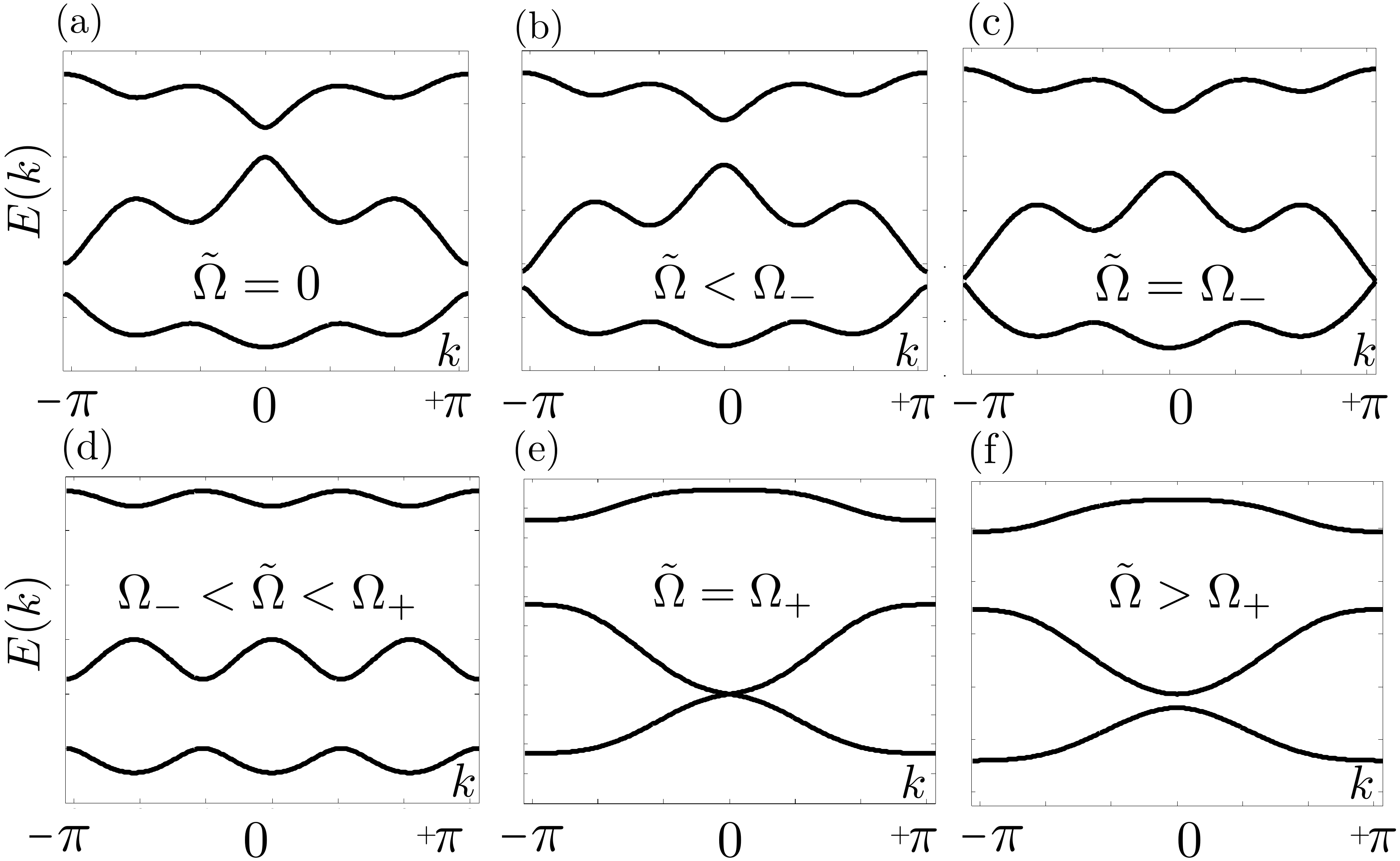}
	\end{center}
	\caption{The spectrum $E(k)$ of the  Hamiltonian $\hat H$ for a three-leg ladder with periodic boundary conditions along the real direction pierced by a flux 
	$\gamma=2\pi/	3$  for different values of the coupling $\tilde \Omega$; $\Omega=1$ and $L_x \rightarrow +\infty$.}
	\label{fig:spectrum}
\end{figure} 

\begin{figure}
	\begin{center}
  	\includegraphics[width=\columnwidth]{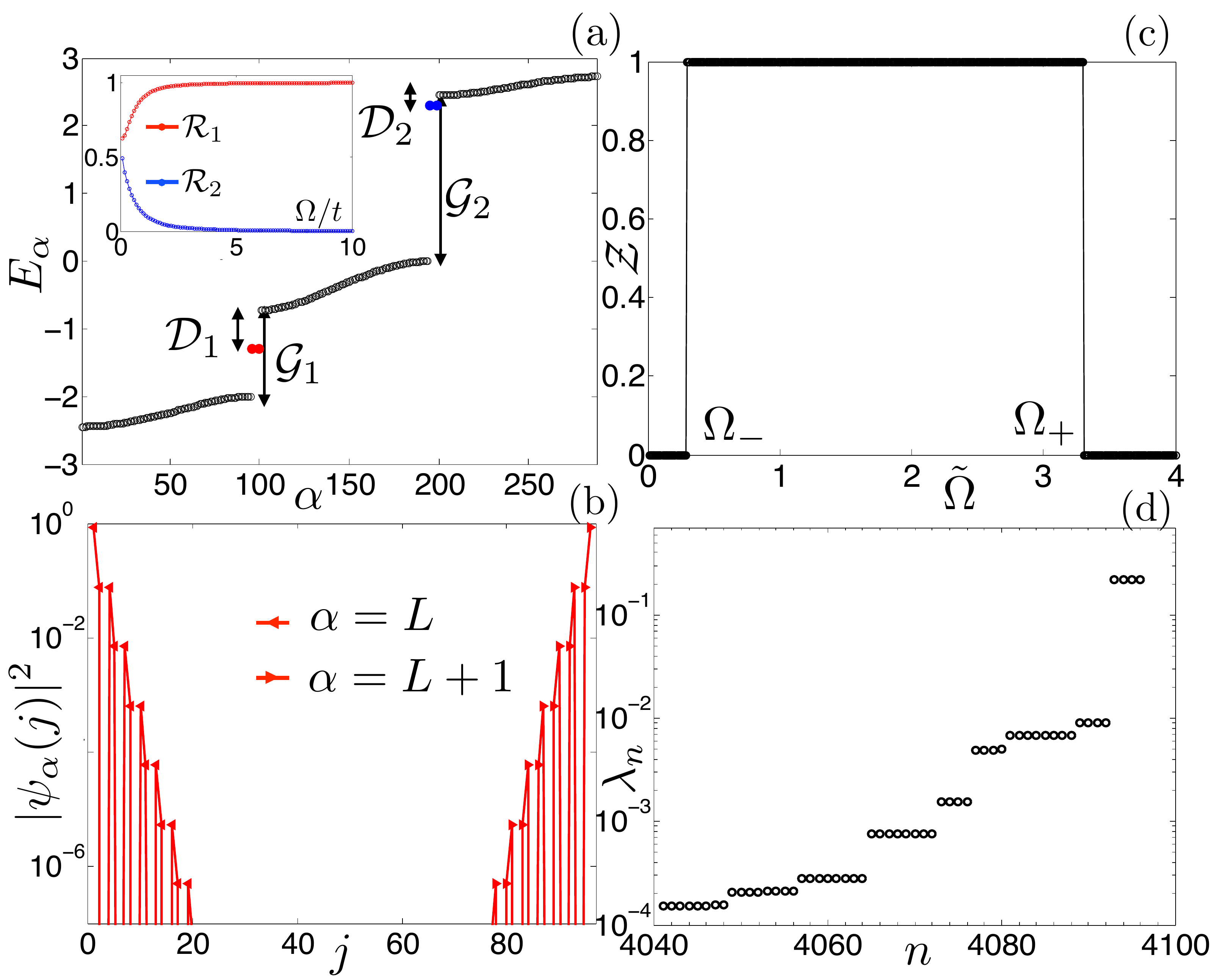}
	\end{center}
	\caption{(a) The eigenvalues $E_\alpha$ of the Hamiltonian $\hat H$ with open boundary conditions along the real direction; (b) the wave-functions corresponding to the two states between the lower and the intermediate band; (c) the Zak phase of the lower band as a function of $\tilde \Omega$; (d) the entanglement spectrum for $\ell=4$;  (a-d) $L_x=96$, 		$L_y=3$, $\gamma=2\pi/3$ and (a-b,d) $\tilde \Omega =\Omega$. }
	\label{fig:topological_evidences}
\end{figure}
As a paradigmatic example, we start considering a three-leg ladder whose phase diagram is shown in Fig.~\ref{fig:model}(b).  To figure out the existence of 
a topologically trivial and a topologically non-trivial phase, we assume periodic boundary conditions along the real direction and 
calculate the spectrum $E(k)$ of the Hamiltonian $\hat H$, see Fig.~\ref{fig:spectrum},  for different values of $\tilde \Omega$ keeping fixed all other parameters; here $\gamma=2\pi/3$. The gap between the lower and the intermediate band  closes for $\tilde \Omega= \Omega_{\pm} =(\pm 3+ \sqrt{4\Omega^2+9})/2 $, suggesting  that, for $ \Omega_-<\tilde \Omega< \Omega_+$, the system is in a topological insulating phase when the lower band is completely filled. 

In order to substantiate our claim, in Fig.~\ref{fig:topological_evidences}(a) we list the eigenvalues $E_\alpha$ ($\alpha=1,\,\dots, 3L_x$) of the Hamiltonian $\hat H$ with open boundary conditions along the real direction and $\tilde \Omega =\Omega$ such that $ \Omega_-<\tilde \Omega< \Omega_+$.  
The existence of two degenerate fermionic states between the lower and the intermediate band when $\Omega_-<\tilde \Omega< \Omega_+$, which disappear if $\tilde \Omega <\Omega_-$ or $\tilde \Omega >\Omega_+$ (not shown), and whose  wave-functions $\psi_\alpha(j)$ are exponentially localized at the edges of the system,  
is indeed a hallmark of a topological phase, see  Fig.~\ref{fig:topological_evidences}(b).

The topological phase is also characterized by a non-local order parameter: the Zak phase~\cite{Zak89,Xiao10} of the lower band $\mathcal{Z}=i/2\pi \int_{-\pi}^{+\pi} \,dk\, \langle u_k|\partial_{k}|u_k \rangle $, which is the Berry's phase picked up by a particle moving across the Brillouin zone, with $u_k(j)=e^{-ikj} \psi_k(j)$ and  $\psi_k(j)$ the Bloch wave-function. Indeed,  in the topological region $ \Omega_-<\tilde \Omega< \Omega_+$, the Zak phase is equal to one and vanishes outside, as shown in Fig.~\ref{fig:topological_evidences}(c).

The existence of a topological phase in our setup can be further proved in terms of the properties of the ground state entanglement spectrum. 
To define this quantity, we consider a subsystem containing $\ell < L_x$ adjacent sites and we call $\overline{\ell}$ its complement; 
then the entanglement spectrum is the set $\{ \lambda_n \}$ of the eigenvalues of the density matrix $\hat \rho_\ell=\mathrm{Tr}_{\overline{\ell}}\left[ \hat \rho \right] $ obtained from the pure density matrix of the ground state of the whole system $\hat \rho= |GS\rangle \langle GS|$. 
It is well known~\cite{Pollmann10,Fidkowski10} that there exists a connection between the topological or trivial nature of the ground state and the degeneracy of the eigenvalues $\{ \lambda_n \}$. A topological phase corresponds to a degenerate entanglement spectrum: this is indeed what we observe in Fig.~\ref{fig:topological_evidences}(d) where the entanglement spectrum is plotted for $\tilde \Omega=\Omega$; outside the topological region, the entanglement spectrum is non degenerate (not shown).

Before going on, it is worth noticing the existence of two additional degenerate edge states between the intermediate and the upper band, see Fig.~\ref{fig:topological_evidences}(a).  Even though they appear when the system is in the topological region, these two states are not protected by a robust gap. In order to elucidate this point, in the inset of Fig.~\ref{fig:topological_evidences}(a),  we plot  the quantity $\mathcal{R}_2= 2\mathcal{D}_2/\mathcal{G}_2$, which measures the ratio between  the minimum gap $\mathcal{D}_2$ of these two states and the upper band, and  the gap $\mathcal{G}_2$ of the upper and the intermediate band; in the same way, we also define the ratio $\mathcal{R}_1$ for the two states between the lower and the intermediate band. 
Indeed we observe that, while $\mathcal{R}_1$ saturates to a finite value inside the topological region when $\Omega$ is increased with $\tilde \Omega=\Omega$; the ratio $\mathcal{R}_2$ goes to zero, thus signaling that the two states between the intermediate and the upper band  are adiabatically connected to the upper band. In light of this, in the following, we focus on the two states between the lower and the intermediate band only. 

\paragraph*{Protecting symmetry.}

\begin{figure}
	\begin{center}
  	\includegraphics[width=\columnwidth]{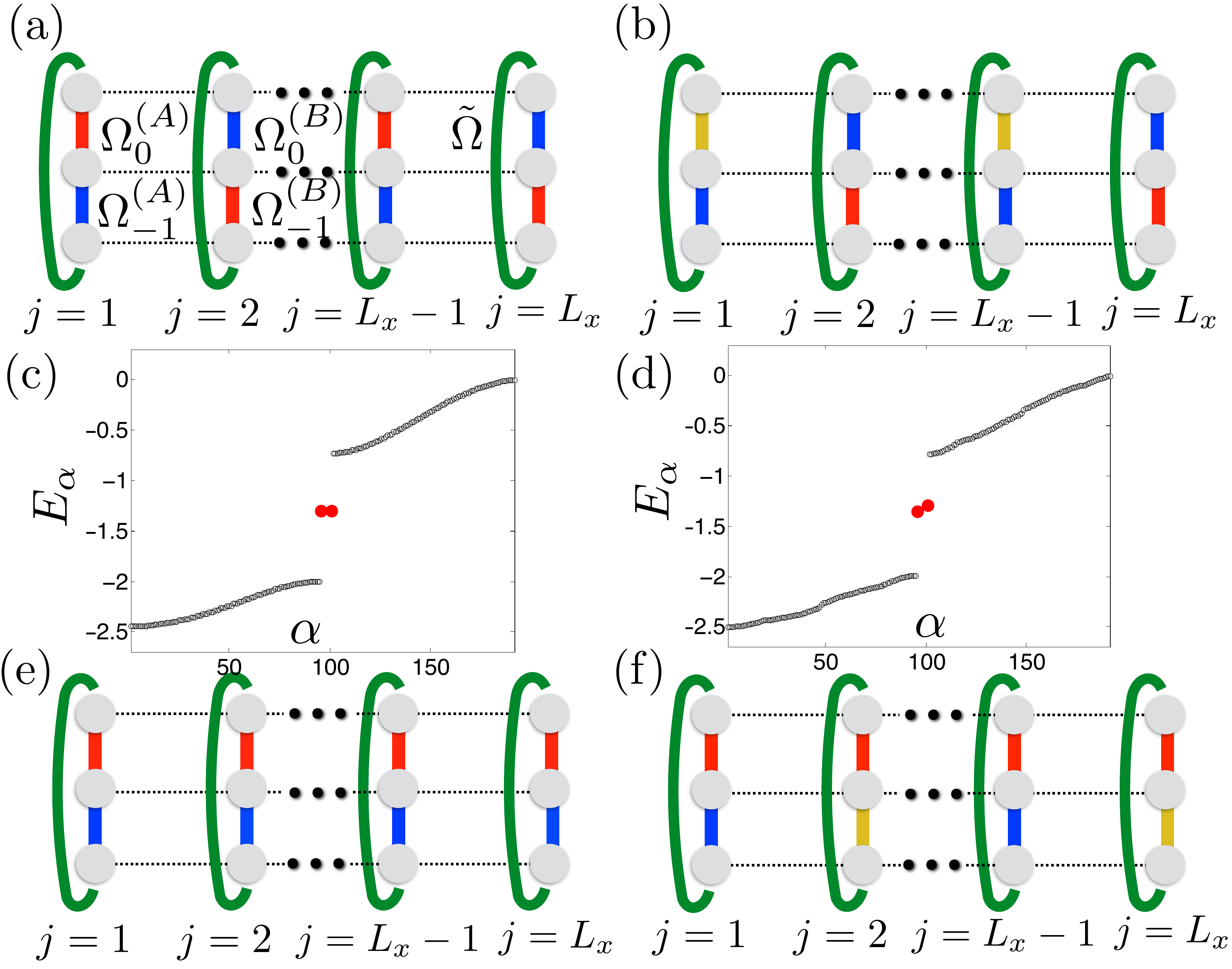}
	\end{center}
	\caption{(a) Schematic plot of a three-leg ladder with $\hat H^{\rm osc}_2$ in the presence (a) and absence (b) of inversion symmetry; 
	eigenvalues with open boundary conditions along the real direction in the presence (c) and absence (d) of inversion symmetry;
	a three-leg ladder with $\hat H^{\rm osc}_2$  in the presence (e) and absence (f) of the generalized anti-unitary inversion symmetry.
	}
	\label{fig:inversion_symmetry}
\end{figure}

We now show that the underlying symmetry protecting the topological phase discussed so far is an inversion symmetry operator $\I$ which generally transforms fermionic operators as
 \begin{equation}
 	\I \, \D_{j,m} \, \I^{-1} =M_{m,m'} \D_{-j,m'},
	\label{inversion_symmetry}
\end{equation}
with $M$ a unitary matrix acting on the internal spin indexes.
To this aim, we need to specify the matrix $M$ such that $\I$ is a symmetry for $\Ham$, \textit{i.e.} $\I \, \Ham \, \I^{-1}= \Ham$, and then we study what happens when $\I$ is broken.

A straightforward calculation shows that the inversion operator $\I_1$ obtained by taking the matrix $M$ in Eq.~\eqref{inversion_symmetry} the anti-diagonal matrix with all entries equal to one, \textit{i.e.} $\I_1 \, \D_{j,m} \, \I^{-1}_1 =\D_{-j,-m}$ is a symmetry of the Hamiltonian $\hat H$ since $\I_1 \, \Ham \, \I^{-1}_1= \Ham$. We also observe the existence of a more subtle inversion symmetry $\I_2$ which acts onto the fermionic operators as $\I_2 \, \D_{j,m} \, \I^{-1}_2 =\D_{-j,m}$. Indeed, $\hat H$ is not invariant under $\I_2$. However, if we take advantage of the anti-unitary time-reversal operator $\T$, \textit{i.e.} $\T \, i \, \T^{-1} = -i$, which acts onto the fermionic operators $\D_{j,m}$ as $\T \, \D_{j,m} \, \T^{-1} = \D_{j,m}$ leaving the internal states $m$ unaltered, it is possible to consider an effective inversion symmetry operator $\I_{\mathcal{T}} \equiv \I_2 \T$ which turns out to be a symmetry of $\Ham$, \textit{i.e.} $\I_{\mathcal{T}} \, \Ham \, \I^{-1}_{\mathcal{T}}= \Ham$. 

In order to break $\I_1$ or $\I_{\mathcal{T}}$, we need to consider a spatial dependent Hamiltonian term in our model.  A convenient way to do that goes through the introduction of the Hamiltonian
\begin{align}
	&\hat H^{\rm osc}_y= \sum_{j }^{\rm odd} \left( \sum_m \Omega^{(A)}_{m} \, \D^\dagger_{j,m} \D_{j,m+1} + \tilde\Omega \, \D_{j,+I}^\dagger \D_{j,-I}  \right)+\nonumber\\
	&+\sum_{j}^{\rm even} \left( \sum_m \Omega^{(B)}_{m} \, \D^\dagger_{j,m} \D_{j,m+1} + \tilde\Omega \, \D_{j,+I}^\dagger \D_{j,-I}  \right)+ \mathrm{H.c.} 
	\label{H2}
\end{align}
which is $\hat H_y$ with spin dependent couplings  and with a  two-site spatial periodicity (what we are going to discuss in the following is valid even if more complicated periodicities are considered). 

The inversion symmetric requirement $\I_1 \, \Ham^{\rm osc}_y \, \I^{-1}_1= \Ham^{\rm osc}_y$ implies $\Omega^{(A)}_{0}=\Omega^{(B)}_{-1}$, and $\Omega^{(A)}_{-1}=\Omega^{(B)}_{0}$, as shown in Fig.~\ref{fig:inversion_symmetry}(a).  When these constraints are satisfied, the two states between the lower and the intermediate band remain degenerate, see Fig.~\ref{fig:inversion_symmetry}(c); when one of them is violated,  \textit{e.g.} by choosing $\Omega^{(A)}_{0} \neq \Omega^{(B)}_{-1}$ as shown in Fig.~\ref{fig:inversion_symmetry}(b), their degeneracy is removed, see Fig.~\ref{fig:inversion_symmetry}(d). This indeed proves that $\I_1$ is a symmetry protecting the topology of the system.  

Let us now examine the role of  $\I_\mathcal{T}$ which is a symmetry of $\Ham^{\rm osc}_y$, \textit{i.e.} $\I_\mathcal{T} \, \Ham^{\rm osc}_y \, \I^{-1}_\mathcal{T}= \Ham^{\rm osc}_y$, 
if $\Omega^{(A)}_{0}=\Omega^{(B)}_{0}$, and $\Omega^{(A)}_{-1}=\Omega^{(B)}_{-1}$, as highlighted in Fig.~\ref{fig:inversion_symmetry}(e). Again, it is sufficient to take $\Omega^{(A)}_{-1} \neq \Omega^{(B)}_{-1}$, as shown in Fig.~\ref{fig:inversion_symmetry}(e), to lift the degeneracy of the two states (not shown). 

To conclude, we stress that the two edge modes between the lower and the intermediate band are protected by two different symmetries  $\I_1$ and $\I_{\mathcal{T}}$. 
In the special case where the couplings $\Omega^{(j)}_{m}$ are not spatial dependent,  $\I_1$ implies $\Omega_{-1}=\Omega_{0}$, while $\I_{\mathcal{T}}$ is satisfied even if $\Omega_{-1} \neq\Omega_{0}$ and, in that context, $\I_{\mathcal{T}}$ generalizes $\I_1$. 

\paragraph*{Topological phase for $L_y>3$.} 

Before discussing the existence of topological phases in synthetic ladders with a number of legs $L_y>3$, 
 it is  worth clarifying that in the regime where inversion symmetry protected topological phases are expected to appear, \textit{i.e.} the flux  per plaquette is $\gamma=2\pi/L_y$, the ladder does not correspond to a two-dimensional geometry even when $L_y \gg 3$. This can be checked by studying the decay of the correlator $|\langle \D^\dagger_{j,m} \D_{j,-I}\rangle|$ along the synthetic dimension which is not purely exponential; see Supplementary Material. 

\begin{figure}
	\begin{center}
 	\includegraphics[width=\columnwidth]{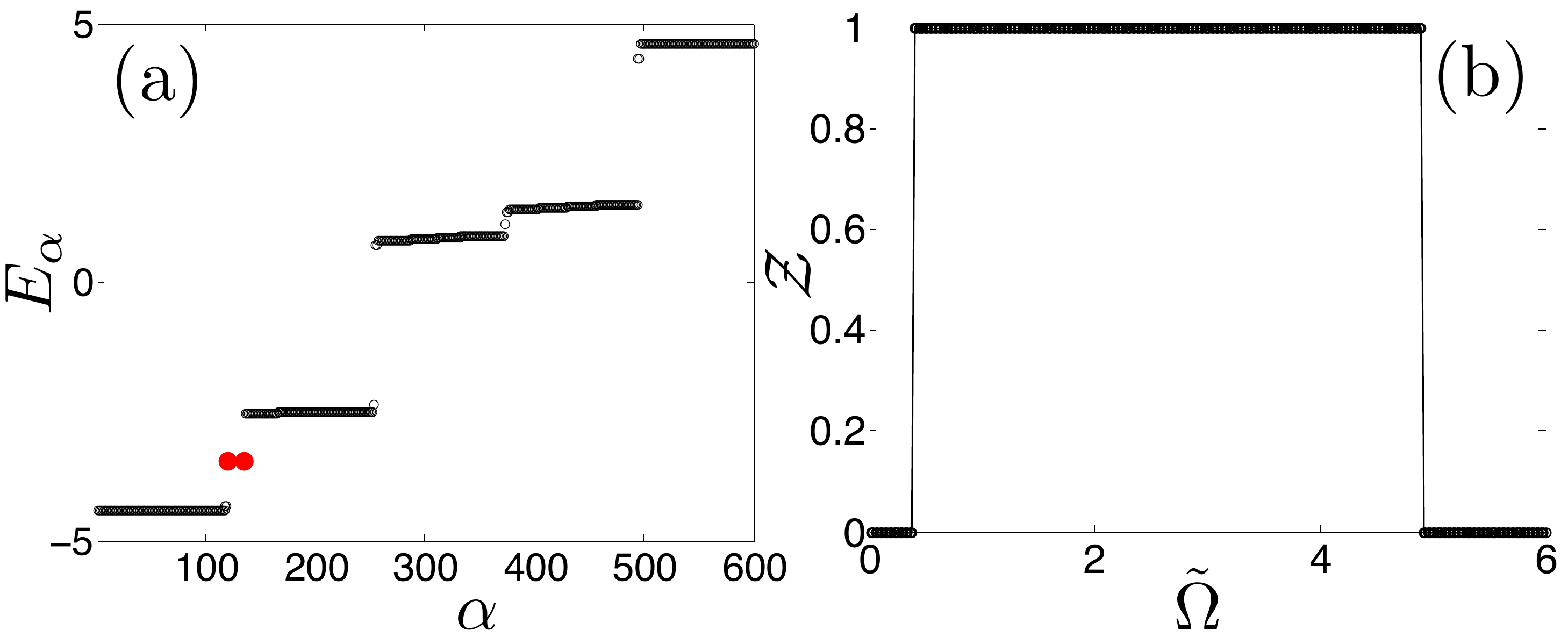}
	\end{center}
	\caption{(a) The eigenvalues $E_\alpha$ of the Hamiltonian $\hat H$ with open boundary conditions along the real direction and $\tilde \Omega=\Omega$; (b) the Zak phase of the lower band as a function 	of $\tilde \Omega$; here $L_x=120$, $L_y=5$, $\gamma=2\pi/5$ and $\Omega=2$.
	}
	\label{fig:su5}
\end{figure}

We now consider the cases odd (even) number of legs $L_y$ separately.
For $L_y$ even and $\gamma=2\pi/L_y$, a closing-opening gap mechanism analogue to the one discussed for $L_y=3$ does not exist. Indeed, the system can be only a trivial band insulator when the particle filling is properly chosen.  
On the contrary, when the synthetic dimension length $L_y$ is odd and $\gamma=2\pi/L_y$, there exists a closing-opening mechanism of the gap similar to the one discussed for $L_y=3$. 
The emerging topological phase is characterized by the appearance of degenerate edge modes, see Fig.~\ref{fig:su5}(a), and by a non-trivial Zak phase, see Fig.~\ref{fig:su5}(b);  again this phase is protected by the inversion symmetry.

\paragraph*{Conclusion.}
We have discussed the appearance of non-interacting topological insulating phases stabilized by the Hofstadter Hamiltonian in a strongly anisotropic ladder  geometry. Protected by a spatial inversion symmetry, these topological phases exhibit fermionic edge modes and can be realized in one-dimensional optical lattices with a synthetic dimension (indeed, these phases are robust against harmonic confinement which leaves the system inversion symmetric, see Supplementary Material).

In this work, we have neglected the role of atom-atom interactions which are expected to drive synthetic ladders pierced by a synthetic magnetic flux into exotic phases  connected with the fractional quantum Hall physics~\cite{Barbarino15, Cornfeld15, Barbarino16, Taddia16, Strinati17,Petrescu16}. Motivated by these previous findings, we leave as an intriguing perspective  a characterization of the topological phases discussed in this paper in the presence of interactions and the search for other interacting topological phases at lower fillings. 

We gratefully thank L. Mazza for his insightful comments. We acknowledge fruitful discussions with L. Privitera, and L. Taddia. 
RF kindly acknowledges support from EU through project QUIC.
GES acknowledges support from EU FP7 under ERC-MODPHYSFRICT.


\clearpage
\begin{center}
{\LARGE Supplementary material}
\end{center}

\begin{center}
{\Large \bf Topological phases in odd-legs frustrated synthetic  ladders}
\end{center}

\begin{center}
{S. Barbarino, M. Dalmonte, R. Fazio, G. E. Santoro}
\end{center}

\setcounter{equation}{0}
\setcounter{figure}{0}
\renewcommand{\figurename}{SUPPL. FIG.}

\subsection*{Derivation of the Hamiltonian $\hat H$}
We consider a one-dimensional chain of length $L_x$ with fermionic atoms endowed with $2I+1$ spin states    
\begin{align}
	\Hamm_1= &\sum_{j=1}^{L_x} \sum_{m=-I}^{I} \left( -t \, \C^\dagger_{j+1,m} \C_{j,m} +  \mathrm{H.c.} \right) +\nonumber\\
	+&\sum_{j=1}^{L_x} \sum_{m=-I}^{I-1} \left(  \Omega \, e^{-i \gamma j}\, \C^\dagger_{j,m} \C_{j,m+1} +  \mathrm{H.c.} \right)	 
	\tag{S1} 
	\label{S1}
\end{align}
where the canonical operator $\C^{(\dagger)}_{j,m}$ annihilates (creates) a fermion at site $j=1,\dots, L_x$ with spin index $m=-I,\dots, I$; $I$  integer (half-integer) corresponds to an odd (even) number of spin states.  
Within the synthetic dimension framework, the Hamiltonian~\eqref{S1} can be interpreted as the Hofstadter  Hamiltonian for spinless fermions on a $L_y=(2I+1)$-leg ladder. The tunneling amplitudes $-t \,e^{i \gamma m}$ between neighboring sites along the real direction
generate the synthetic magnetic field corresponding to a uniform  flux $+\gamma$ per plaquette; $\Omega$ is the coupling strength  between adjacent spin states along the synthetic direction. 
We also assume that the extremal spin states $-I$ and $+I$ are coupled by
\begin{align}
	\Hamm_2= \sum_{j=1}^{L_x} \left(  \tilde \Omega \, e^{-i \gamma j}\, \C^\dagger_{j,I} \C_{j,-I} +  \mathrm{H.c.} \right) \, ;
	 \tag{S2} 
\end{align}
within the synthetic dimension framework this last terms converts the ladder into a cylinder pierced by a flux $+\gamma$ per plaquette.

If we apply the unitary transformation $\hat U \C_{j,m} \hat U^\dagger = e^{-im \gamma j} \D_{j,m}$  onto $\Hamm_1+\Hamm_2$, \textit{i.e.}
$\hat H= \hat U (\Hamm_1+\Hamm_2) \hat U^\dagger$,  we obtain 
\begin{align}
	&\hat H= \sum_{j=1}^{L_x} \sum_{m=-I}^{+I} \left( -t \, e^{i \gamma m}\, \D^\dagger_{j+1,m} \D_{j,m} + \mathrm{H.c.} \right) + \nonumber\\
	&+\sum_{j=1}^{L_x} \left(  \sum_{m=-I}^{+I} \Omega \,  \D^\dagger_{j,m} \D_{j,m+1} + \tilde\Omega_j \, \D_{j,+I} \D_{j,-I} + \mathrm{H.c.}  \right)\, ,
	\tag{S3}  \label{S2}
\end{align}
with $\tilde \Omega_j \equiv \tilde \Omega \, e^{-i\gamma (2I+1) j}$, which is indeed the Hamiltonian considered in the main text.

\subsection*{Derivation of $\Omega_\pm$ }
We start diagonalizing the Hamiltonian~\eqref{S2} with $I=1$ and $\gamma=2\pi/3$ and periodic boundary conditions along the real direction. We conveniently introduce the momentum space operators $\D_{k,m}= 1/\sqrt{L_x} \sum_j e^{ikj} \D_{j,m}$ with $k \in [-L_x/2,\dots, L_x/2-1]$ and we define $\hat D^\dagger_{k} = \left(\D^\dagger_{k,-1} \; \; \D^\dagger_{k,0} \;\; \D^\dagger_{k,+1} \right)$ such that
$
\hat H = \sum_{k} \hat D^\dagger_{k} H_k \hat D_{k}
$
with
\begin{equation}
	H_k= \left( \begin{matrix}  -2t \cos \left(k-\frac{2\pi}{3}\right)& \Omega&\tilde \Omega \\ \Omega&-2t \cos k&\Omega \\ \tilde \Omega &0&-2t \cos \left(k+\frac{2\pi}{3}\right) \end{matrix} \right)
	\tag{S4}
\end{equation}
and we solve the secular equation $\mathrm{det}[H_k-E\, \mathrm{Id} ]=0$, with $\mathrm{Id}$ the identity matrix ($\Omega, \, \tilde \Omega$ are assumed to be real) which gives
\begin{align}
&E^3-(3t^2+2\Omega^2-\tilde \Omega^2)E+2t^3 \cos 3k +\nonumber\\
&-2t (\tilde \Omega^2-\Omega^2) \cos k - 2\Omega^2 \tilde \Omega=0\,.
\tag{S5} \label{S4}
\end{align}
Even though Eq.~\eqref{S4} admits an analytical solution for the three bands
$E_\eta(k)$, with $\eta=L \mathrm{(lower)}, \; I \mathrm{(intermediate)}, U \mathrm{(upper)}$, we observe that  the values of $\Omega_\pm$ can obtained by taking into account that the gap closes at $k=\pi$ (for $\Omega_-$) and at $k=0$ (for $\Omega_+$). Then, to derive $\Omega_-$, it is sufficient to require $E_L(k=\pi)=E_I(k=\pi)$ which gives
\begin{align}
\Omega_- = \frac{1}{2} \left(-3t+\sqrt{4\Omega^2+9t^2} \right)
\tag{S6} \label{S5}
\end{align}
Similarly, $\Omega_+ = 1/2 \left(3t+\sqrt{4\Omega^2+9t^2} \right)$ is obtained by requiring $E_L(k=0)=E_I(k=0)$.

\subsection*{Charge fractionalization}
\begin{figure}
	\begin{center}
  	\includegraphics[width=0.8\columnwidth]{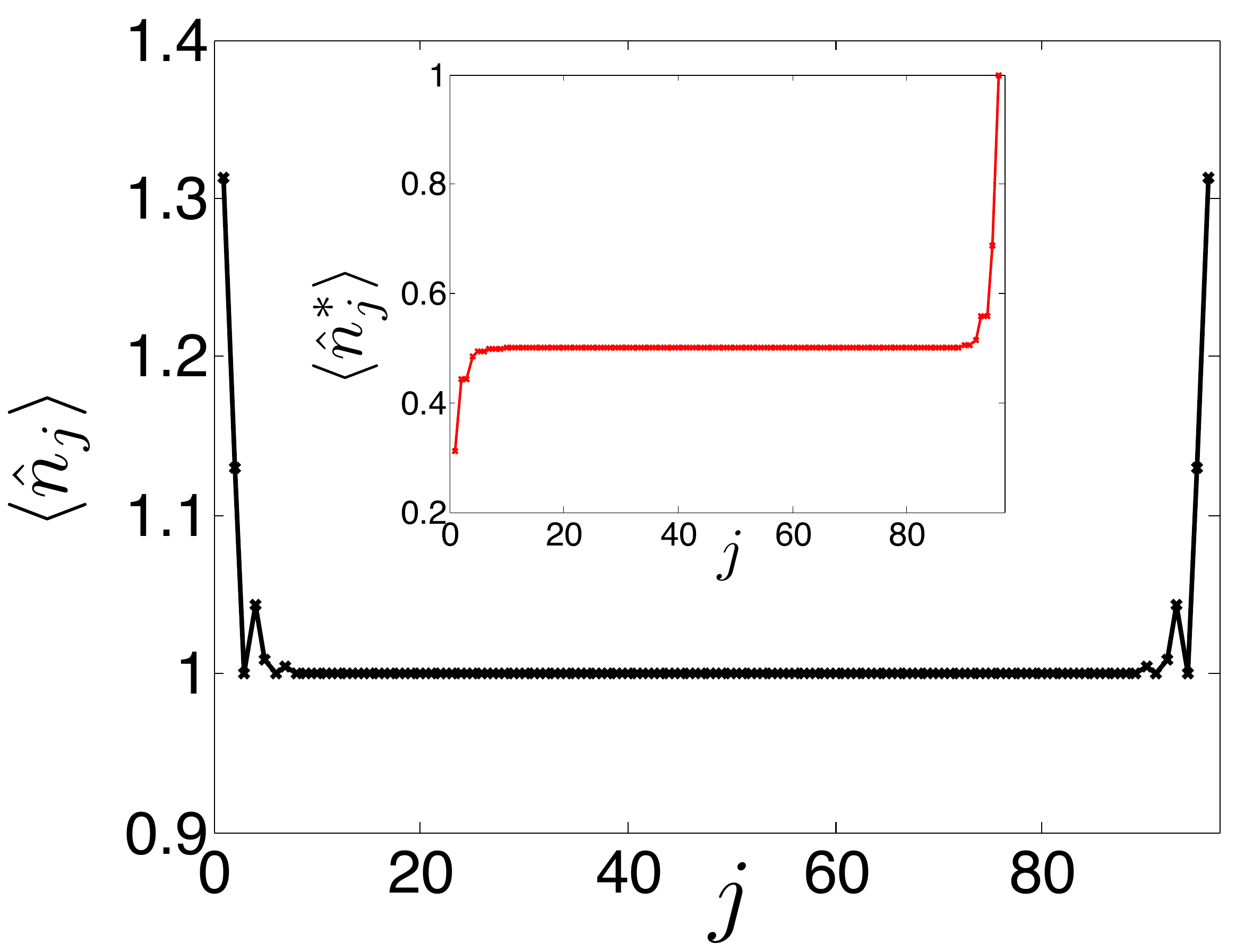}
	\end{center}
	\caption{The density profile for a three-leg ladder of length $L_x=96$ sites and $N=97$ particles; here $\Omega=\tilde \Omega=1$ and $\gamma=2\pi/3$. 
	The inset displays the expectation value of $ \hat n^*_j $. 
	}
	\label{suppl_fig2}
	\end{figure}

In the topological phase, the presence of the two fermionic modes exponentially localized at the edges leads to a striking density profile when $N=L_x+1$ particles are present. 
Indeed, if we plot the expectation value of the density operator
\begin{equation} \tag{S7}
	\hat n_j=\sum_{m=-1}^{+1} \,  \D^\dagger_{j,m} \D_{j,m} \; ,
\end{equation}
we observe that it is flat in the bulk, while half a particle is localized at each boundary, as shown in Suppl. Fig.~\ref{suppl_fig2}. This important signature of particle fractionalization can be identified 
through the operator~\cite{Mazza15}
\begin{equation} \tag{S8}
	\hat n^*_j=\sum_{\ell=1}^j \left(\hat n_\ell-1 \right) 
\end{equation}
which is shown in the inset of Suppl. Fig.~\ref{suppl_fig2}.

\begin{figure}
	\begin{center}
  	\includegraphics[width=\columnwidth]{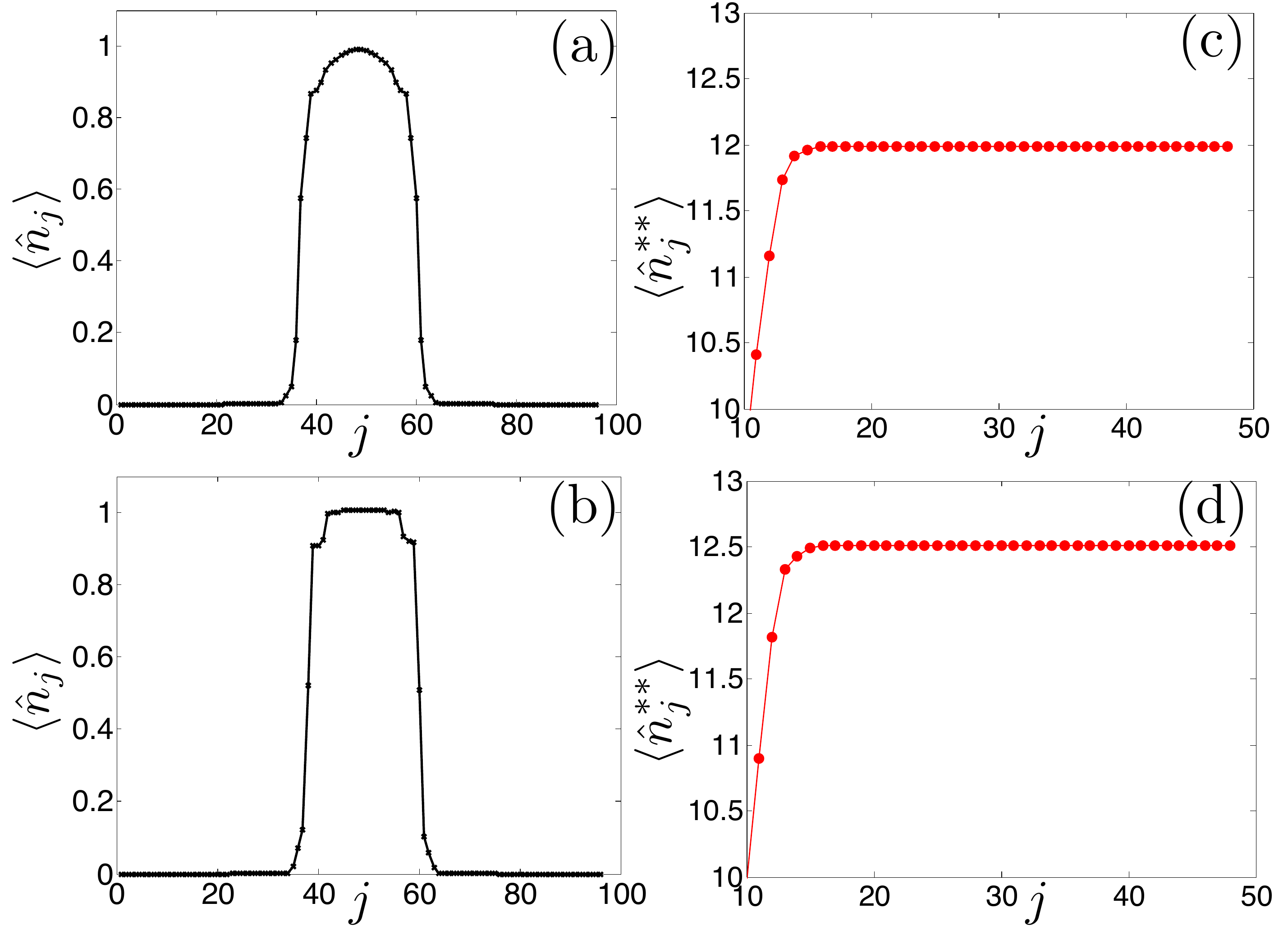}
	\end{center}
	\caption{Density profile and the expectation value of the operator $\hat n^{**}_j$ in the presence of harmonic confinment in the trivial (a,b) and in the topological region (b,d); $L_x=96$, 
	$\Omega=1$, $\omega=0.006$ and $N=22$ particles; $\tilde \Omega=0$ in the trivial region and $\tilde \Omega=\Omega$ in topological region. 
	}
	\label{suppl_fig3}
	\end{figure}

Particle fractionalization can be observed even in the presence of a harmonic confinement 
\begin{equation} \tag{S9}
	\hat H_{\rm trap} = \sum_{j,m} \omega_{j} \, \D^\dagger_{j,m} \D_{j,m} 
\end{equation}
with $\omega_j=\omega (j-\frac{L_x}{2})(j-\frac{L_x}{2}-1)$. To this aim, following Ref.~\cite{Mazza15}, we introduce the operator
\begin{equation} \tag{S10}
	\hat n^{**}_j=\sum_{\ell=0}^j \left(\hat n_{L_x/2+\ell}-1 \right) \,.
\end{equation}
In Suppl. Fig.~\ref{suppl_fig3}(a,c) we plot the density profile both in the trivial and in the topological region and
the corresponding expectation value of $\hat n^{**}_j$, see Suppl. Fig.~\ref{suppl_fig3}(b,d).
 
\subsection*{Disorder}
\begin{figure}
	\begin{center}
  	\includegraphics[width=\columnwidth]{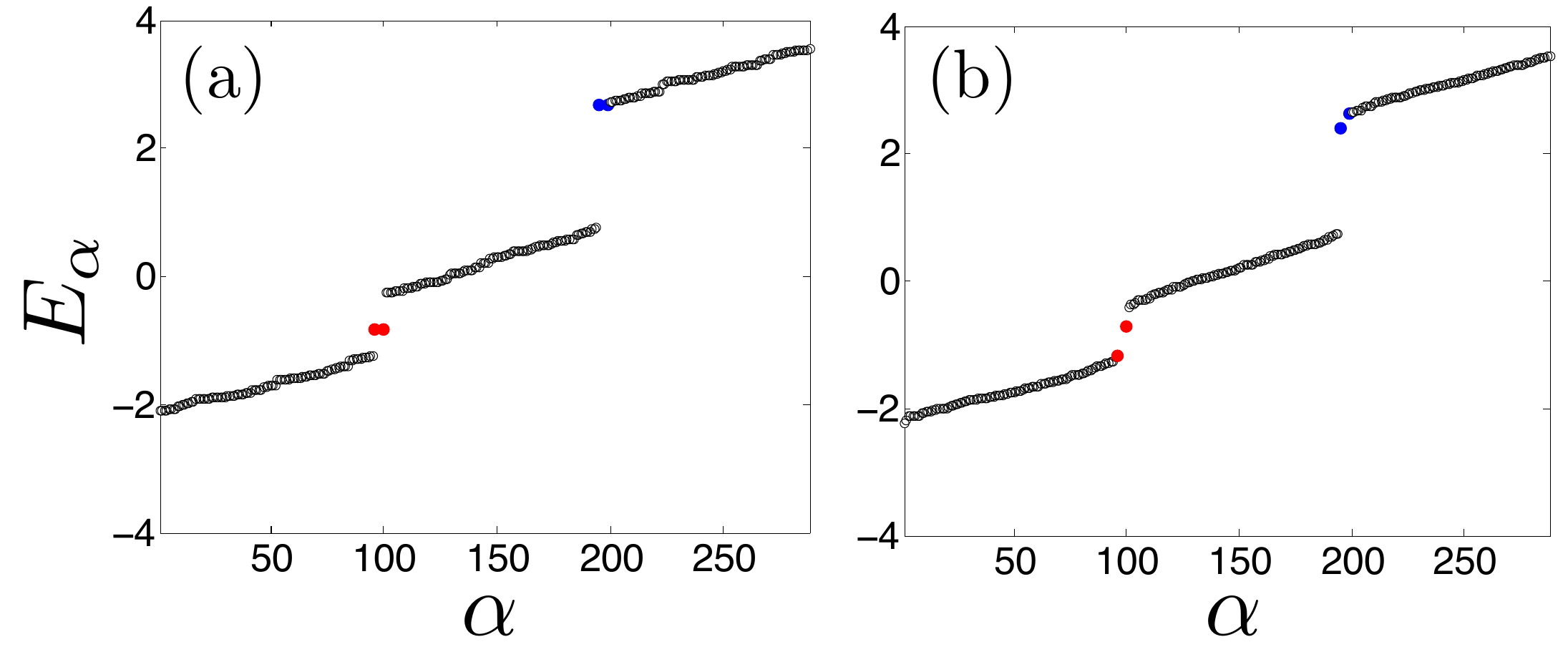}
	\end{center}
	\caption{The eigenvalues $E_\alpha$ of the Hamiltonian $\hat H + \hat H_{\rm noise}$ in the presence of inversion symmetric disorder (a), completely random disorder (b);
	here $L_x=96$, $\Omega=\tilde \Omega=1$ and $\gamma=2\pi/3$. 
	}
	\label{suppl_fig4}
	\end{figure}

We study the faith of the fermionic edge modes in a three-leg ladder in the presence of random impurities described by the Hamiltonian
\begin{equation} \tag{S11}
	\hat H_{\rm noise}= \sum_{j=1}^{L_x} \sum_{m=-1}^{+1}\, \delta V_j \, \D^\dagger_{j,m} \D_{j,m} \, .
\end{equation}
As shown in Suppl. Fig.~\ref{suppl_fig4}(a), when the disorder is inversion symmetric, \textit{i.e.} $\delta V_j=\delta V_{L_x+1-j}$, the degeneracy of the fermionic edge states is unaltered, 
on the contrary,  it is removed by a completely random disorder, see Suppl. Fig.~\ref{suppl_fig4}(b).

\subsection*{Correlator}

\begin{figure}
	\begin{center}
  	\includegraphics[width=\columnwidth]{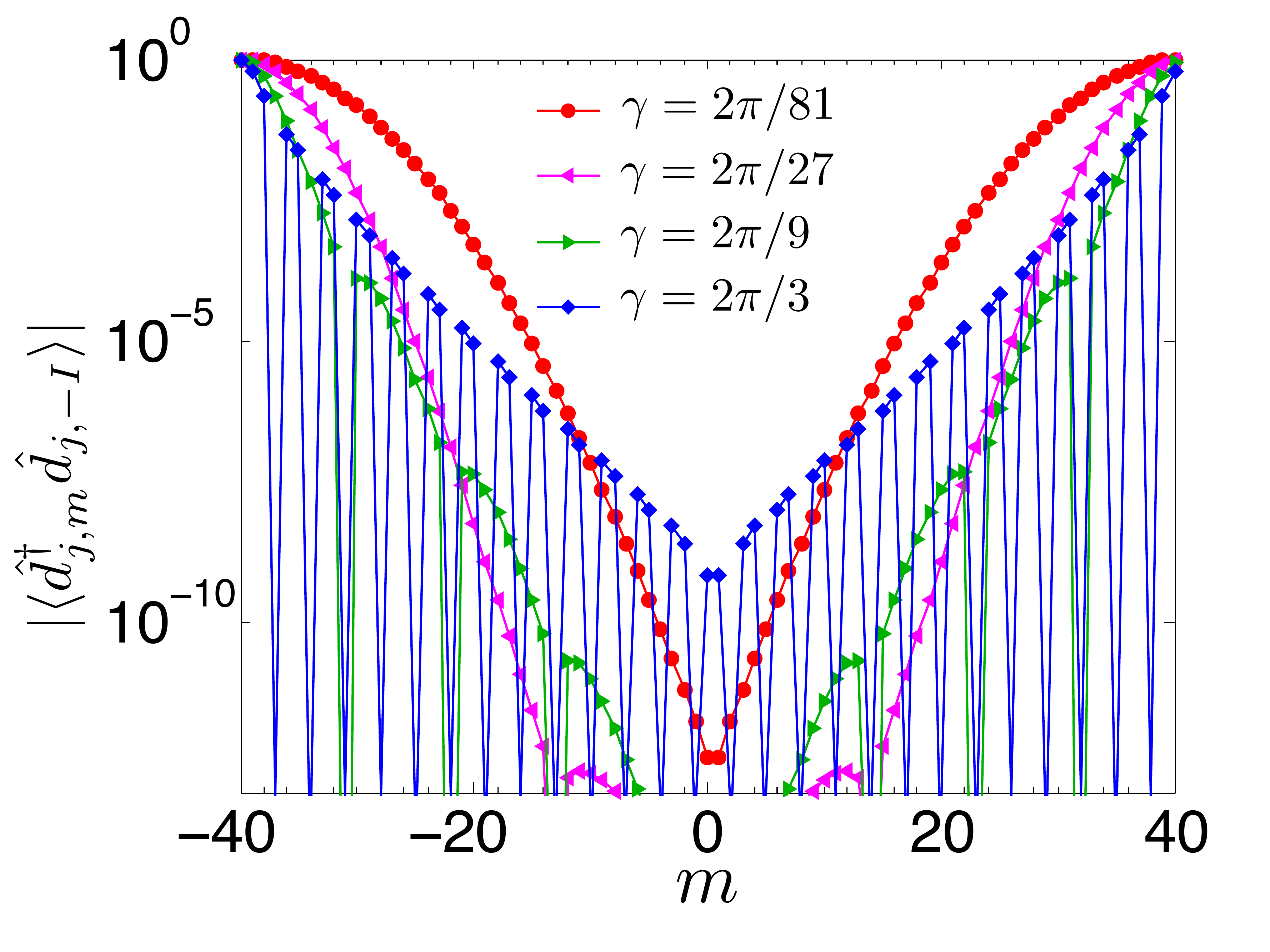}
	\end{center}
	\caption{The correlator $|\langle \D^\dagger_{j,m} \D_{j,-I}\rangle |$ as a function of $m$ for a $L_y=81$-leg ladder and 
 	different fluxes $\gamma$; $\Omega=\tilde \Omega=1$. Periodic boundary conditions both along the real and the synthetic directions are used and  $L_x \gg L_y$.
	}
	\label{suppl_fig}
	\end{figure}

In order to understand if a cylinder with transverse dimension $L_y$  describes a two-dimensional or a one-dimensional theory, we study the correlator $|\langle \D^\dagger_{j,m} \D_{j,-I}\rangle|$ along the synthetic direction in $L_y=81$-leg ladder  (periodic boundary conditions along both directions are assumed and the particle filling is tuned such that the ground state is gapped; $\langle \; \rangle$ indicates the ground state average), see Suppl. Fig.~\ref{suppl_fig}.
Indeed, when the flux per plaquette is large, \textit{e.g.} $\gamma=2\pi/3$, the correlator decays exponentially on a length scale much smaller than $L_y$ and thus the ladder describes the physics of the Hofstadter Hamiltonian in two dimensions. On the contrary, when the flux $\gamma$ per plaquette is tuned to $\gamma=2\pi/81$, \textit{i.e.} the value for which the system enters the inversion symmetry protected topological phase, the correlator $|\langle \D^\dagger_{j,m} \D_{j,-I}\rangle|$ does not decay with a pure exponential behavior signaling that the ladder does not describe a two-dimensional theory anymore. 
Summarizing, in the regime where topological phases are expected to appear, \textit{i.e.} when the flux per plaquette is $\gamma=2\pi /q$ with $q \sim L_y$, the physics is not two-dimensional even if $L_y \gg 1$ and this is even more true if  $L_y$ is of the order of few sites such as in synthetic ladders.

\end{document}